\documentclass[a4paper,11pt]{article} % option doublespacing or reviewcopy for double line spacing

\usepackage{graphicx} % for more complicated commands
\usepackage{amsmath,amsfonts,amssymb,bm}
\usepackage{color}

\usepackage[utf8]{inputenc}

\usepackage{microtype} % for ligatures  ff fl fi, etc
\DisableLigatures{encoding = *, family = *}

\usepackage{ragged2e} % has command \justifying

\usepackage{fancyhdr}
\fancypagestyle{plain}{\pagestyle{fancy}} % adds on the first page as well

\usepackage{blindtext}

\usepackage[sort&compress,numbers]{natbib}
\usepackage{doi}%<----------
\begin{document}

\title{Recovering almost everything diffusion could reveal}

%\justifying

\author{Evren \"Ozarslan\\ %, Second Authos$^b$  \\
	{}\footnotesize\it Department of Biomedical Engineering, Campus US, Linköping University, SE-581 85 Linköping, Sweden. \\
\footnotesize\it  E-mail:evren.ozarslan@liu.se} %\\
%	{}$^b$ \normalsize\it Department of Neuroscience, University of	Florida, Gainesville, FL 32610, USA.  }
\date{}

%\footnote{E-mail:evren.ozarslan@liu.se } 

\maketitle
\pagestyle{fancy}

\begin{abstract}
Diffusion magnetic resonance has been employed for determining the distribution of net displacements (ensemble average propagator), moments and correlations of net displacements, and the steady-state distribution of magnetized particles. All such quantities are accessible via the diffusion propagator, which characterizes the diffusion process fully. Here, a novel diffusion encoding and data analysis framework is introduced with which the diffusion propagator can be recovered.
\end{abstract}

\section{Introduction}

The diffusion propagator $P(\mathbf{x}',t|\mathbf{x})$ represents the probability that a particle located at position $\mathbf{x}$ travels to $\mathbf{x}'$ over a time interval of duration $t$. This quantity fully describes the diffusive motion in simple as well as complicated environments, which could feature reflective, relaxing, or semi-permeable walls, external forces, spatially varying diffusivity, etc. Such structural parameters imprint their trace on the propagator. The propagator, if available, could reveal a detailed static picture of the space within which diffusion is taking place. 

\begin{figure}[t!]
	\begin{center}
	\includegraphics[width=.55\textwidth]{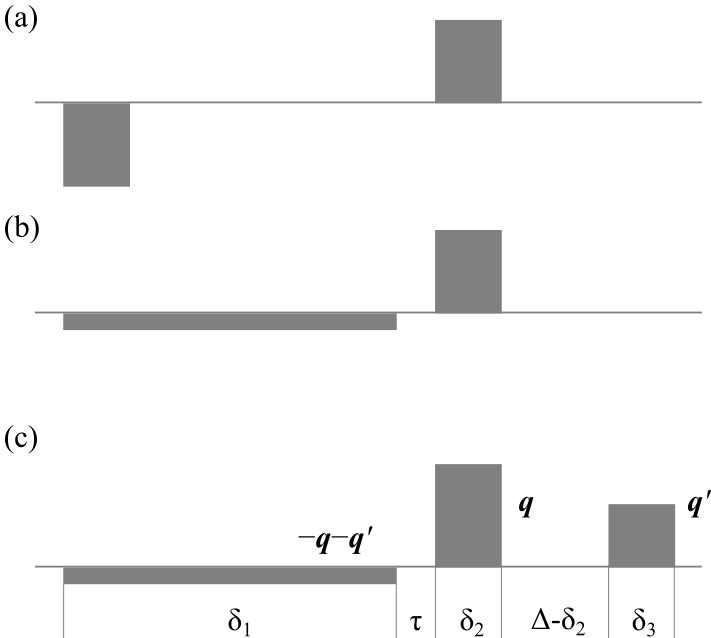}
\end{center}
	\vspace{-14pt}\caption{(a) Effective gradient waveform of the Stejskal-Tanner sequence \cite{StejskalTanner65}. The signal can be transformed into the ensemble average propagator \cite{Stejskal65}. (b) Effective gradient waveform of the sequence by Laun et al \cite{Laun11prl}, which can be utilized to obtain the long-time form of the diffusion propagator. (c) Effective waveform for one realization of the experiment considered. Here, $-\mathbf q-\mathbf q’$, $\mathbf q$, and $\mathbf q’$ are the signed areas under the first, second, and third gradient pulses, respectively. The signal can be transformed into the actual propagator.  \label{fig:pulse}}
\end{figure}

For example, consider diffusion within a domain bounded by partially relaxing walls. If the diffusion propagator is available, the bulk diffusivity can be determined from its short-time behavior; the domain shape is given by its support; and the surface relaxivity can be obtained from its decay, most conveniently at long times.
The most commonly employed scheme for diffusion encoding is that introduced by Stejskal and Tanner\cite{StejskalTanner65} in 1965, see Figure \ref{fig:pulse}a. The MR signal attenuation for this sequence has been related to an ensemble average propagator (EAP) since its early days\cite{Stejskal65}. The EAP represents the distribution of net displacements, which is the diffusion propagator substantially compromised due to an integration over all possible initial positions.

More recently, another two-pulse sequence has been introduced \cite{Laun11prl} (see Figure \ref{fig:pulse}b) featuring one long pulse compensating another narrower pulse. This method provides a means to determine the pore geometry, or more adequately, the steady-state distribution of the particles, which is just the long-time limit of the diffusion propagator.

Here, a new experimental and data-analysis framework is introduced, which reveals the actual diffusion propagator.

\section{Methods and Results}

\subsection{Pulse sequence and theory}

One realization of the effective waveform proposed in this work is depicted in Figure \ref{fig:pulse}c. The sequence comprises two independently-determined narrower gradient pulses, whose collective integral is compensated by one long pulse. Within a $d$-dimensional pore space, and in the ideal scenario of infinitely long first pulse followed by infinitesimally narrow pulses, the detected MR signal is simply the Fourier transform  
$$E_\Delta(\mathbf{q},\mathbf{q}')=\int \mathrm{d} \mathbf{x} \, \rho(\mathbf{x}) \int \mathrm{d}\mathbf{x}' \, P(\mathbf{x}', \Delta | \mathbf{x}) \, e^{-i (\mathbf{q}\cdot\mathbf{x}+\mathbf{q}'\cdot\mathbf{x}' )}\ ,$$
allowing the determination of the propagator via inverse Fourier transforms through the expression
$$P(\mathbf{x}',\Delta|\mathbf{x}) =\frac{\int\mathrm{d}\mathbf{q}\, e^{i\mathbf{q}\cdot\mathbf{x}}\int\mathrm{d}\mathbf{q}'\, e^{i\mathbf{q}'\cdot\mathbf{x}'}\, E_\Delta(\mathbf{q},\mathbf{q}')}{(2\pi)^d \int\mathrm{d}\mathbf{q}\,e^{i\mathbf{q}\cdot\mathbf{x}}\, E_\Delta(\mathbf{q},\mathbf{0})}\ .$$

\begin{figure}[h!]
	\begin{center}
		\includegraphics[width=.61\textwidth]{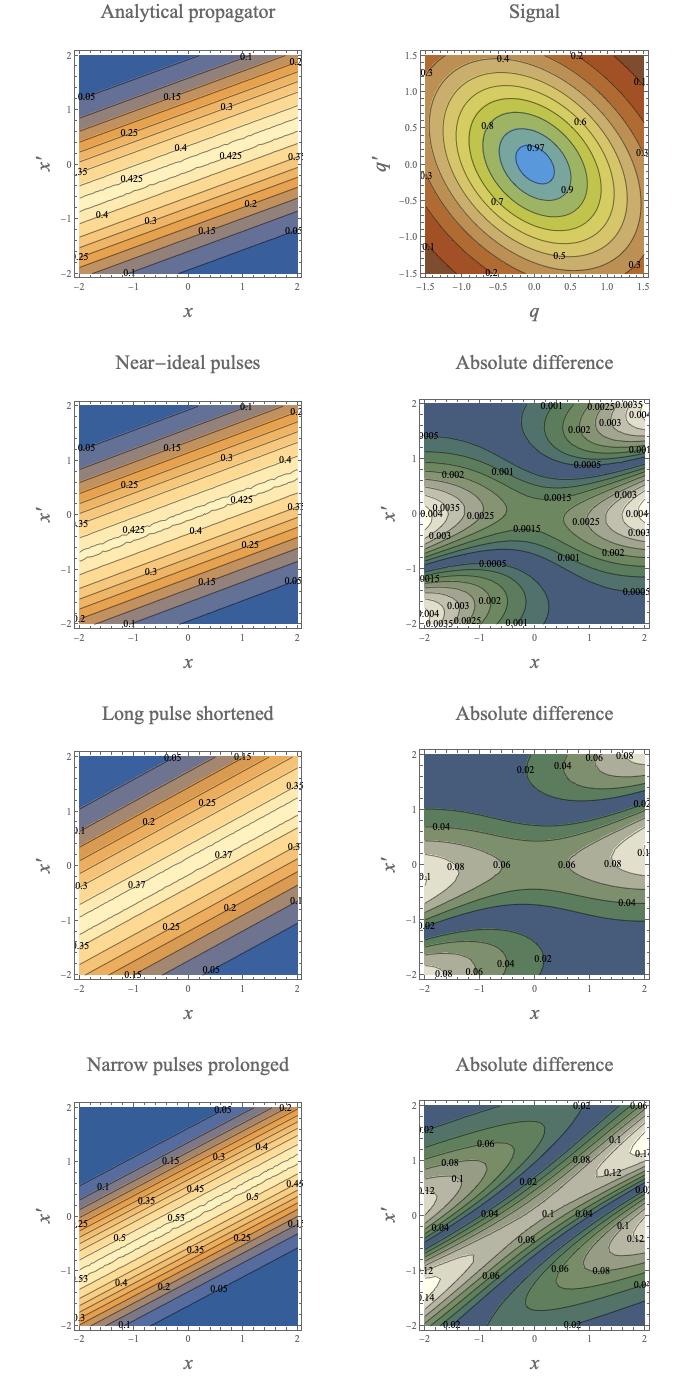}
	\end{center}
	\vspace{-24pt}\caption{ Top row: Contour plots for the analytical propagator and the associated `ideal' signal. Second row: estimated propagator and its deviation from the analytical one for near-ideal parameters: $\delta_1=100, \tau=0.1, \delta_2=\delta_3=0.002$, and $\Delta=1$. Third row: same with $T=1$. Fourth row: same with  $\delta_2=\delta_3=0.9$. All quantities are expressed in multiples of the characteristic time ($\beta^{-1}$) and length scales ($(D/\beta)^{1/2}$) of the process, where $\beta$ is the force constant divided by the friction coefficient. \label{fig:OUprocess}}
\end{figure}

\subsection{Validation and beyond}

The technique was validated for the problem involving one-dimensional diffusion near an attractive center \cite{Stejskal65} for which the analytical expression for the propagator is available \cite{Uhlenbeck30}. The signal and its transform via the above expression were evaluated for arbitrary parameters. It was shown to yield the analytical form of the propagator for ideal pulses, i.e., in the limit $\delta_1\rightarrow \infty$, $\delta_2\rightarrow 0$, and $\delta_3 \rightarrow 0$.

Figure \ref{fig:OUprocess} shows a summary of our results. As the duration of the long pulse is shortened, the propagator obtained through the above procedure is blurred. On the other hand, the finite width of the narrow pulses changes the meaning of the propagator such that the displacements are defined between the centers-of-mass of the trajectories traversed during the application of the pulses \cite{MitraHalperin95}.

A signature of the presence of external forces is the violation of the reciprocity condition, i.e., $P(\mathbf{x}',t|\mathbf{x})\neq P(\mathbf{x},t|\mathbf{x}’)$, which is revealed by the asymmetry of the estimated propagator about the $\mathbf{x}=\mathbf{x}'$ line.

\subsection{Exchanging restricted compartments}

\begin{figure}[h!]
	\begin{center}
	\includegraphics[width=.99\textwidth]{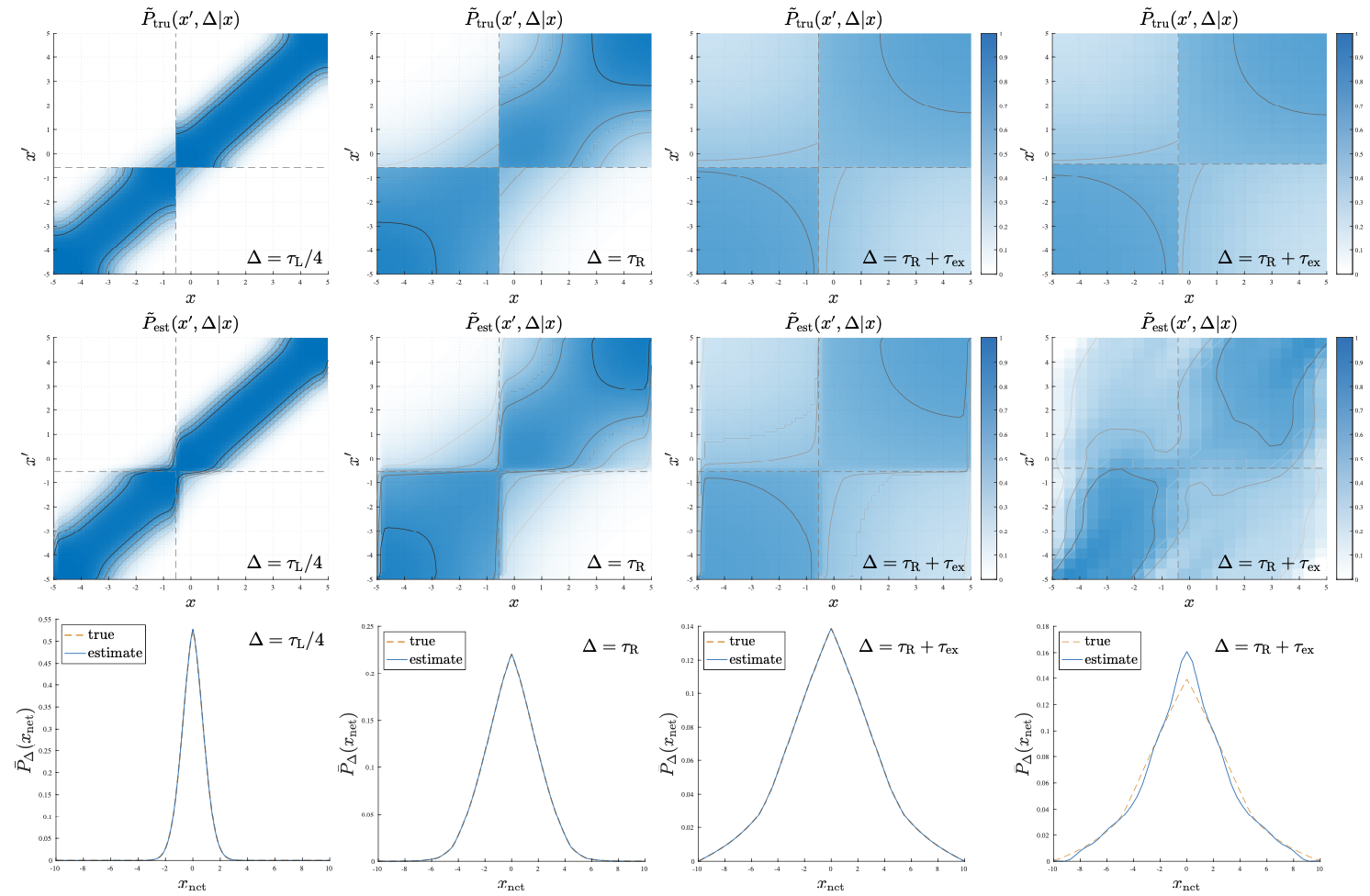}
	\end{center}
	\vspace{-20pt}\caption{Simulations for two compartments separated by a membrane of permeability ${w}=0.6\,\mu$m/ms. Compartment sizes: $L_\mathrm{L}=4.5\,\mu$m (left), $L_\mathrm{R}=5.5\,\mu$m (right), diffusivities: $D_\mathrm{L}=3\,\mu$m$^2/$ms, $D_\mathrm{R}=2\,\mu$m$^2/$ms. Time-scales: $\tau_\mathrm{L}=L_L^2/\pi^2D_L$, $\tau_\mathrm{R}=L_R^2/\pi^2D_R$, $\tau_\mathrm{ex}=\sqrt{D_\mathrm{L} D_\mathrm{R}}/{w}^2$. Top to bottom: true propagator, estimated propagator, EAP. Left three columns: near-ideal parameters. Last column parameters: $\delta_1=200\,$ms, $\delta_{2,3}=1.2\,$ms, and $G_\mathrm{max}=10\, $T/m. \label{fig:permeable}}
\end{figure}

\begin{figure}[h!]
	\begin{center}
		\includegraphics[width=.49\textwidth]{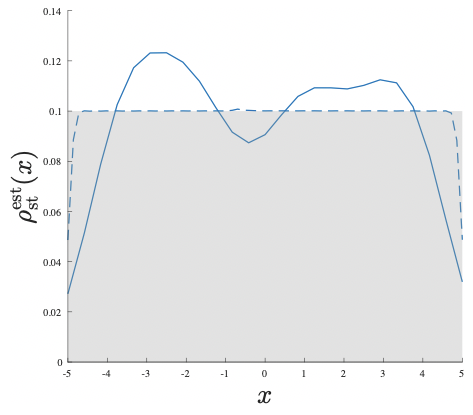}
	\end{center}
	\vspace{-24pt}\caption{Estimate of the steady state density for the experimental parameters of the right column of Figure 3 (solid line) and the near-ideal parameters (dashed line). The shaded rectangle indicates the true uniform steady state density of $1/L$. \label{fig:GSest}}
\end{figure}

By repeating the experiment with different values of $\Delta$, the method allows the study of diffusive dynamics. In Figure \ref{fig:permeable}, we illustrate our results for diffusion within and between two restricted compartments featuring different diffusivities as well as sizes separated by a thin semi-permeable membrane. For this one-dimensional problem, the propagator can once again be displayed on a plane for each diffusion time (first three columns). At short times, the diffusivities can be estimated. In the intermediate time point considered, the molecules spread within their compartment and start leaking into the neighboring one. At longer times, a significant portion of the molecules end up in the other compartment. In the last column of Figure \ref{fig:permeable}, we show the results for parameters that can be employed in some preclinical scanners. The method provides interpretable results elucidating the structure within the pore space in great clarity. This is in sharp contrast to the case of EAP (bottom row) and the steady-state distribution (Figure \ref{fig:GSest}), which are non-specific even for ideal experimental conditions. 

\section{Discussion and Conclusion}

We introduced a new pulse sequence featuring two independent gradients and related the signal to the diffusion propagator through an expression involving Fourier transforms. Note that our method reduces to the Stejskal-Tanner sequence for $\mathbf{q}'=-\mathbf{q}$, and to Laun et al.'s sequence for $\mathbf{q}'=\mathbf{0}$. Our technique should not be confused with position exchange spectroscopy \cite{Han00}, which also employs two independent pulses to encode positions making it suitable for flow measurements. For diffusion within a heterogeneous specimen composed of many pores, a long pulse is introduced to prevent large cancellations of signals for pores located at different positions. 

The diffusion propagator is the key quantity that fully describes diffusion for Markov processes. Compared to EAP, the true diffusion propagator provides exquisite sensitivity to the fine characteristics of the pore space because of the extra dimension made available by our technique. As such, the method could be a powerful probe that could transform our capabilities of imaging tissue microstructure.

\section*{Acknowledgments}

After conceptualizing the technique, the author had discussions with Cem Yolcu, who performed the simulations in Figure \ref{fig:permeable}, Carl-Fredrik Westin, and Magnus Herberthson. Nicolas Moutal and Denis Grebenkov provided the code for their work (J Sci Comput, 81, 3, p. 1630-1654, 2019), which was helpful for the simulations in Figure \ref{fig:permeable}.

\bibliography{\string~/Dropbox/SharedBib/sharedbib}

\end{document}